\documentclass{iopart}

\usepackage{amssymb,graphicx}
\usepackage{epsfig}

\begin{document}

\title{Robustness of the Blandford-Znajek mechanism}

\author{Carlos Palenzuela$^{1,2}$,
Carles Bona$^{3}$,
Luis Lehner$^{2,4,5}$,
Oscar Reula$^{2,6}$}
\address{
$^{1}$Canadian Institute for Theoretical Astrophysics, Toronto, Ontario M5S 3H8,
 Canada.}
\address{
$^{2}$Department of Physics \& Astronomy, Louisiana State University, Baton Rouge, LA 70802, USA.}
\address{
$^{3}$Institute for Applied Computing with Community
Code (IAC$^{\,3}$).\\
Universitat de les Illes Balears, Palma de Mallorca 07122, Spain.}
\address{
$^{4}$Perimeter Institute for Theoretical Physics,Waterloo, Ontario N2L 2Y5, 
Canada}
\address{
$^{5}$Department of Physics, University of Guelph, Guelph, Ontario N1G 2W1, 
Canada}
\address{
$^{6}$FaMAF, Universidad Nacional de Cordoba, Ciudad Universitaria, Cordoba 5000, Argentina 
}

\begin{abstract}
The Blandford-Znajek mechanism has long been regarded as a key
ingredient in models attempting to explain powerful jets
in AGNs, quasars, blazzars etc. In such mechanism, energy is extracted
from a rotating black hole and dissipated at a load at far distances. 
In the current work we examine the behaviour of the BZ mechanism with respect to
different boundary conditions, revealing the mechanism robustness upon
variation of these conditions. Consequently, this work closes a gap in our
understanding of this important scenario.
\end{abstract}

\maketitle
\section{Introduction}
The Blandford-Znajek (BZ) effect~\cite{1977MNRAS.179..433B} has been proposed
as one key mechanism to explain energetic jets from black hole systems.
In this mechanism a spinning hole interacts with magnetic fields sourced
by an accretion disk (see. e.g.~\cite{Noble:2008tm,McKinney:2006dy,McKinney:2004ka}). The 
magnetic field extracts rotational energy
from the spinning black hole and dissipates this energy at a far away load. 
The details of this process are complex and its understanding
requires analyzing the behaviour of the plasma interacting with the strong curvature region
around the black holes and how the energy extracted, which is represented by a powerful
collimated Poynting flux, is dissipated at large distances.

A simple picture explaining this process is provided by the membrane 
paradigm~\cite{Thorne:1986iy}, where the system
is modeled as a circuit composed of a battery --provided by the charge separation induced on the black hole--,
two long wires --along magnetic field lines-- and a load --at large distances from the black hole--.
The energy tapped from the spinning black hole is understood in terms of an induced EMF flowing
from the pole to the equator and released at the load. While this picture provides a basic understanding
of the underlying phenomena, the role of the load in the resulting jet and energy dissipation has not been
the subject of a detailed analysis. In particular, the possibility that different loads could significantly affect the jet has not been explored.
If such were the case, this mechanism could be questioned as a robust ingredient to explain black hole systems producing collimated energy outputs. 

In this note, in order to investigate this issue, we examine the influence of the 
boundary conditions on the dynamics of the system
and, in particular, on the resulting Poynting flux energy. By considering different conditions, 
playing the role of the load, we illustrate the robustness of this mechanism. Our results
indicate that while these conditions do result in some differences, these do not significantly affect  the jet structure and energetics of the system.

\section{Physical system}
The system is modeled by assuming the force free approximation in a curved
background provided by a Kerr black hole spacetime. The black hole is initially
embedded in a pure magnetic field configuration, with its intrinsic angular momentum
(or spin) aligned with the asymptotic magnetic field.
We study the dynamics of the electromagnetic fields numerically and examine the induced
Poynting flux. The details of our formulation and implementation are provided
in~\cite{2010arXiv1005.1067P,2010arXiv1007.1198P}. We here consider more general 
boundary conditions, which will allow us to examine scenarios
corresponding to different resistive loads.
To define such conditions we must first understand the characteristic structure of the system

\subsection{Characteristic decomposition}

Our starting point are the (general relativistic version of the) Maxwell equations. Further, 
we consider the augmented version of the equations, which include the divergence-cleaning fields
$\{ \phi,\Psi \}$ to dynamically control the constraint violations 
\cite{Komissarov:2007,Palenzuela:2009hx}. This system is defined by
\begin{eqnarray}
  (\partial_t - {\cal L}_{\beta}) E^{i} &-& \epsilon^{ijk} \nabla_j (\alpha B_k) 
   + \alpha \gamma^{ij} \nabla_j \Psi = \alpha trK E^i - \alpha J^i \, ,
\label{maxwellext_3+1_eq1a} \\
  (\partial_t - {\cal L}_{\beta}) \Psi &+& \alpha \nabla_i E^i = 
    \alpha q -\alpha \kappa \Psi \, ,
\label{maxwellext_3+1_eq1b} \\
  (\partial_t - {\cal L}_{\beta}) B^{i} &+& \epsilon^{ijk} \nabla_j (\alpha E_k) 
   + \alpha \gamma^{ij} \nabla_j \phi = \alpha trK B^i \, ,
\label{maxwellext_3+1_eq1c} \\
  (\partial_t - {\cal L}_{\beta}) \phi &+& \alpha \nabla_i B^i = 
   -\alpha \kappa \phi \, .
\label{maxwellext_3+1_eq1d}
\end{eqnarray}
It is illustrative to consider first the electrovacuum case (i.e. $J^i=q=0$). The characteristic
structure can be computed by considering the propagation of perturbations along a generic
direction $n^i$ (belonging to an orthonormal tetrad $\{p,q,n\}$, where the index
$n$ stands for the longitudinal component).
We consider an arbitrary solution (labeled generically by $u$) and formulate the
eigenvalue problem for the perturbed fields (labeled by $[u]$). By defining
$\beta^n \equiv \beta^i n_i$ and computing the (Lagrangian) velocities 
${\tilde v} \equiv v + \beta^n$, this eigenvalue problem can be written as
\begin{eqnarray}
  {\tilde v} [E^{i}] &=& -\epsilon^{ijk} n_j [\alpha B_k] + n^i [\alpha \Psi] \, ,
\label{RH_eq2a} \\
  {\tilde v} [B^{i}] &=& \epsilon^{ijk} n_j [\alpha E_k] + n^i [\alpha \phi] \, ,
\label{RH_eq2b} \\
  {\tilde v} [\Psi] &=& \alpha [E^n] \, ,
\label{RH_eq2c} \\
  {\tilde v} [\phi] &=& \alpha [B^n] \, .
\label{RH_eq2d}
\end{eqnarray}
 It is now straightforward to obtain the following list of eigenvectors:
\begin{itemize}
  \item constraint modes: $[\phi] \pm [B^n]$ and $[\Psi] \pm [E^n]$,
         which involve the divergence-cleaning fields $\{ \phi,\Psi \}$ as well as
         the divergences of $E$ and $B$. They propagate with light
         speed  $v=-\beta^n \pm \alpha$.
  \item transversal modes: $[E^i - E^n n^i] \mp [\epsilon^{ijk} n_j B_k]$,
         which can be written in tetrad components as $[E^p] \pm [B^q]$ and $[E^q] \mp [B^p]$.
        They correspond to the EM waves and also propagate with light speed $v=-\beta^n \pm \alpha$.
\end{itemize}
Notice that due to the linearity of the Maxwell equations in vacuum,
the eigenvectors depend only on the perturbed fields. Therefore, at this level,
our discussion is completely generic and independent of the background solution $u$.

Let us now consider a more realistic approach to our physical problem, --a spinning
black hole interacting with a magnetic dominated, low density plasma, surrounding it--.
The effect of this plasma is to provide both a charge $q$ and current distribution $J^i$ which 
profoundly affect the behavior of the electromagnetic fields. As described 
in~\cite{Goldreich:1969sb,1977MNRAS.179..433B}, and summarized next, their role is accounted
by explicitly defining the component of $J^i$ transversal to $B^i$ 
and introducing a further constraint to account for the component of $J^i$ parallel to $B^i$, while
$q$ is obtained through Coulomb's law. 
In the following discussion, 
we will ignore the contribution of the divergence-cleaning terms since we are
interested in describing the physical set-up. In our problem of interest, the plasma
inertia is negligible, consequently the Lorentz force vanishes, defining the so-called force-free 
condition~\cite{Goldreich:1969sb,1977MNRAS.179..433B},
\begin{equation}\label{ff_condition}
     q E^i + \epsilon^{ijk} J_j B_k = 0 ~~.
\end{equation}
This condition determines the current required to close Maxwell equations.
The scalar and vector products of the force-free
condition (\ref{ff_condition}) with the magnetic field lead respectively to $E_i B^i = 0$
and $J^i = q \epsilon^{ijk} E_j B_k/B^2 + (J_i B^i) B^i/B^2$. The component of the current
parallel to the magnetic field can
be obtained by imposing in the Maxwell equations $(\partial_t - {\cal L}_{\beta}) (E_i B^i) =0$,
which is a natural consequence of the condition $E_i B^i = 0$.  Additionally, in this electrodynamic
limit the charge density
is defined through the Maxwell constraint $q = \nabla_i E^i$, so that the current only depends
on the electromagnetic fields,
\begin{eqnarray}\label{current}
\fl
 J^i = 
     (\nabla_m E^m) \epsilon^{ijk} E_j B_k/B^2 
     + B^i\,\epsilon^{jkm}\,(B_j \nabla_k B_m - E_j  \nabla_k E_m )/B^2 \, .
\end{eqnarray}
The force-free evolution system is obtained from the original Maxwell system
(\ref{maxwellext_3+1_eq1a}-\ref{maxwellext_3+1_eq1d}), by substituting current (\ref{current})
in equation (\ref{maxwellext_3+1_eq1a}). Additionally, the field $\Psi$ and its
associated evolution equation (\ref{maxwellext_3+1_eq1b}) are eliminated since they
become trivial with the definition of $q = \nabla_i E^i$. The eigenvalue problem for the 
force-free approximation can be written as
\begin{eqnarray}
\fl  ~~~~~{\tilde v} [E^i] &=& -\alpha \epsilon^{ink} [B_k] + \alpha\, (S^i/B^2) [E_n] +
                     \alpha\, (B^i/B^2) \epsilon^{jnk} (B_j [B_k] - E_j [E_k]) \, ,
\nonumber
\label{RH_eq4a} \\
\fl  ~~~~~{\tilde v} [B^i] &=& \alpha \epsilon^{ink} [E_k] + \alpha n^i [\phi] \, ,
\nonumber
\label{RH_eq4b} \\
\fl  ~~~~~{\tilde v} [\phi] &=& \alpha [B^n] \, ; 
\label{RH_eq4c}
\end{eqnarray}
where $S^i \equiv \epsilon^{ijk} E_j B_k$ is the Poynting vector.
The diagonalization of the system is considerably more involved
now due to the non-linearities introduced by the current (\ref{current}).
A convenient rearrangement of the fields makes this task easier,
by changing to a basis containing the following combinations
\begin{eqnarray}
 &[C]& \equiv B_q[B_p] - B_p[B_q] + E_p[E_q] - E_q[E_p] \, ,
\nonumber \\
 &[E_t^2 + B_t^2]&  \equiv E_p[E_p] + E_q[E_q] + B_p[B_p] + B_q[B_q] \, ,
\nonumber \\
 &[E_i B^i]& = E_n[B_n] + B_n[E_n] + [E_t B^t] \, ,
\nonumber \\
 &[E_t B^t]& \equiv E_p[B_p] + B_p[E_p] + E_q[B_q] + B_q[E_q] \, .
\end{eqnarray}
The characteristic problem in this basis is given by
\begin{equation}
\fl
{\tilde v} \,  {\bf U} = 
\alpha\, \left( \begin{array}{ccccccc}
S_n/B^2 & B_n(B^2-E^2)/B^2 & 0 & 0 & 0 & E_n & -1 \\
B_n/B^2 & S_n/B^2 & 0 & 0 & 0 & 0 & 0 \\
0 & -E_n & 0 & 1 & 0 & 0 & 0 \\
-E_n B_n/B^2 & -S_n E_n/B^2 & 1 & 0 & 0 & 0 & 0 \\
0 & 0& 0 & 0 & 0 & 1 & 0 \\
0 & 0& 0 & 0 & 1 & 0 & 0 \\
0 & 0& 0 & 0 & E_n & 0 & 0 
 \end{array} \right) {\bf U}
\end{equation} 
with
\begin{equation}
{\bf U} = \left[ 
C ~~~ E_n ~~~ S_n ~~~ E_t^2+B_t^2 ~~~ \phi_n ~~~ B_n ~~~ E_i B^i 
 \right]^{{\bf T}} \, .
\end{equation} 
The characteristic matrix can be diagonalized to obtain the following list of eigenvectors
\begin{itemize}
  \item constraint modes: $[\phi] \pm [B_n]$,
        which propagate with light speed
        $v=-\beta^n \pm \alpha$.
  \item standing mode: $[E_i B^i] -E_n [B_n]$,
        which propagates with speed $v=-\beta^n$, and contains information on $[E_t B^t]$.
  \item ``Poynting" modes: $-B_n\,\lambda^{\pm}[C] + (B^2 \lambda^+\,\lambda^-\,
         - S_n\,\lambda^{\pm})[E_n] -E_n\,B_n\,[B_n] + B_n [E_i B^i]$, which propagate with speed 
       $v=-\beta^n + \alpha\,\lambda^{\pm}$, where we have defined
        $\lambda^{\pm} \equiv \frac{S_n}{B^2} \pm \frac{B_n}{B^2} \sqrt{B^2-E^2}$.
  \item transversal radiative modes: $B^2(\lambda^+\,\lambda^- - 1)E_n[E_n] 
        + 2\,B_n\,E_n [E_i B^i] + B^2(\lambda^+\,\lambda^- + 1 \mp 2 S_n) ([E_t^2 + B_t^2] \pm [S_n])
        \mp 2\,E_n\,B_n [C] \pm 2\, E_n^2 B_n [\phi]$, which are a generalization of the
        standard MHD Alfven modes, , 
        and propagate with light speed $v=-\beta^n \pm \alpha$.
\end{itemize}


The force-free evolution system is strongly hyperbolic since there is a complete basis of 
eigenvectors for each direction $n^i$, ensuring the existence of a unique stable solution
within the domain of dependence of the initial hypersurface in a boundary-free case.
If the system is to be employed within a finite domain, boundary conditions have to be imposed on 
(some of) the fields, which may affect the stability of the solution. To ensure
the well posedness of the resulting problem care must be taken to define these fields
consistently. One way to do so for symmetric hyperbolic systems, relies on adopting maximally
dissipative boundary conditions
of the type $U^{-} = R\,U^{+}$, where $U^{\mp}$ are the ingoing/outgoing eigenvectors of the
evolution system, with $|R| < 1$~\cite{gko}. In a problem
 with constraints --as is the case here-- further
conditions must be derived to ensure the constraints are preserved.
In the case of the force-free system as we have shown previously, the characteristic
structure is rather complicated and depends on the background solution which complicates the analysis
\footnote{See e.g.~\cite{Cecere:2008sj}
for a related example in the MHD case where constraint preserving boundary conditions
are defined.}. We defer to a future work the construction
of constraint-preserving boundary conditions for the force-free system of equations.
Here, since our main interest is to model different behaviours of the resistive load far from
the black hole, we will consider the eigenvectors of the
electrovacuum Maxwell system as a first step towards consistent
and stable boundary conditions for the force-free system. Notice that this is short
of defining conditions consistent with a force-free regime, however at far distances from the
black hole, the plasma density decreases reducing to the vacuum case. Thus, depending
on the physical scenario considered this choice will not be a physical limitation. Nevertheless,
regardless of this issue, our approach will allow us to explore the robustness of the BZ mechanism
by considereding different conditions and assesssing the stability and collimation of 
the resulting (if any) Poynting flux.
The boundary conditions considered here take the form
\begin{eqnarray}
    \left( [E^i - E^n n^i] - [\epsilon^{ijk} n_j B_k] \right) &=& R \, 
    \left( [E^i - E^n n^i] + [\epsilon^{ijk} n_j B_k] \right) \, ,\\
    \left( [ \phi]  - [B_n] \right) &=& R \,  \left( [\phi] + [B_n] \right) ~~,~~
    [E_n] = 0 \, .
\end{eqnarray}
Notice that there is no rigorous proof that the resulting system will be stable
with $|R| < 1$ except in the electrovacuum case.
Numerical experiments have shown that it is unstable if $R < 0$ but stable otherwise.

\section{Results}
We adopt the formulation of the force-free approximation and the numerical implementation
already described in~\cite{2010arXiv1005.1067P,2010arXiv1007.1198P}. We will consider
different boundary conditions corresponding to $R=\{0,1/2,1\}$ in order to model the role
of the resistive load far away from the source. Notice that we do not claim any of these values
to be a true representative of the load --which is unknown in any case--, rather these 
values allow us to explore the resulting
behaviour under profoundly different physical conditions and examine, in particular,
the resulting Poynting flux behaviour. Indeed the resulting conditions contemplate that
in absence of divergence errors, $[B_n]=0$, and for:
\begin{itemize}
\item $R=1$: the transversal components of the electric field are free 
(i.e., determined only by the interior solution)
while the transversal components of the magnetic field are set to 0.
\item $R=0$: the transversal components of the magnetic field are equal of the transversal
components of the electric field (and determined by the interior solution). 
\end{itemize}
Any other value in $R \in (0,1)$ just spans the parameter space between these two limits.

\begin{figure}
\begin{center}
\epsfig{file=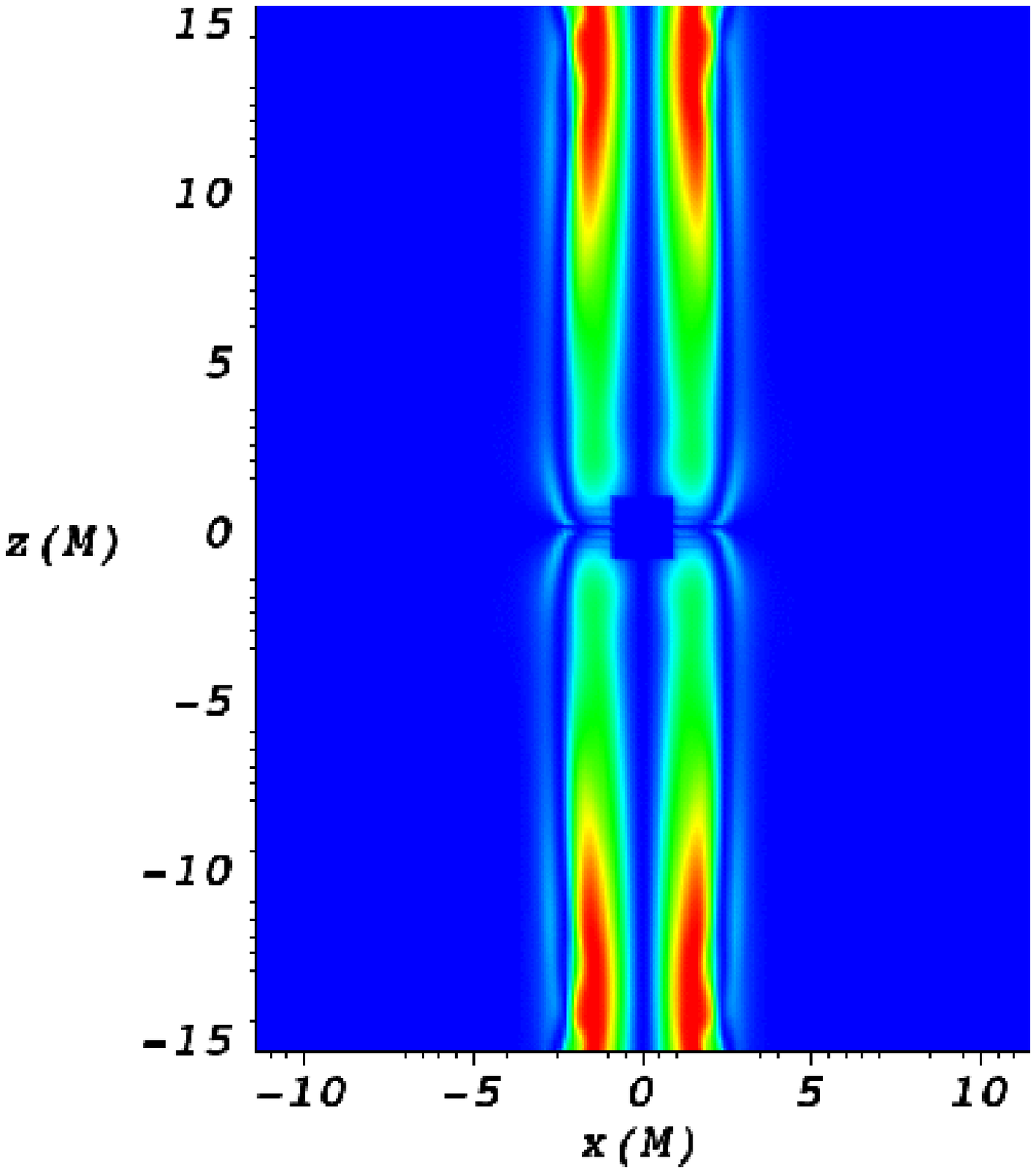, height=5.0cm,width=4.2cm}
\epsfig{file=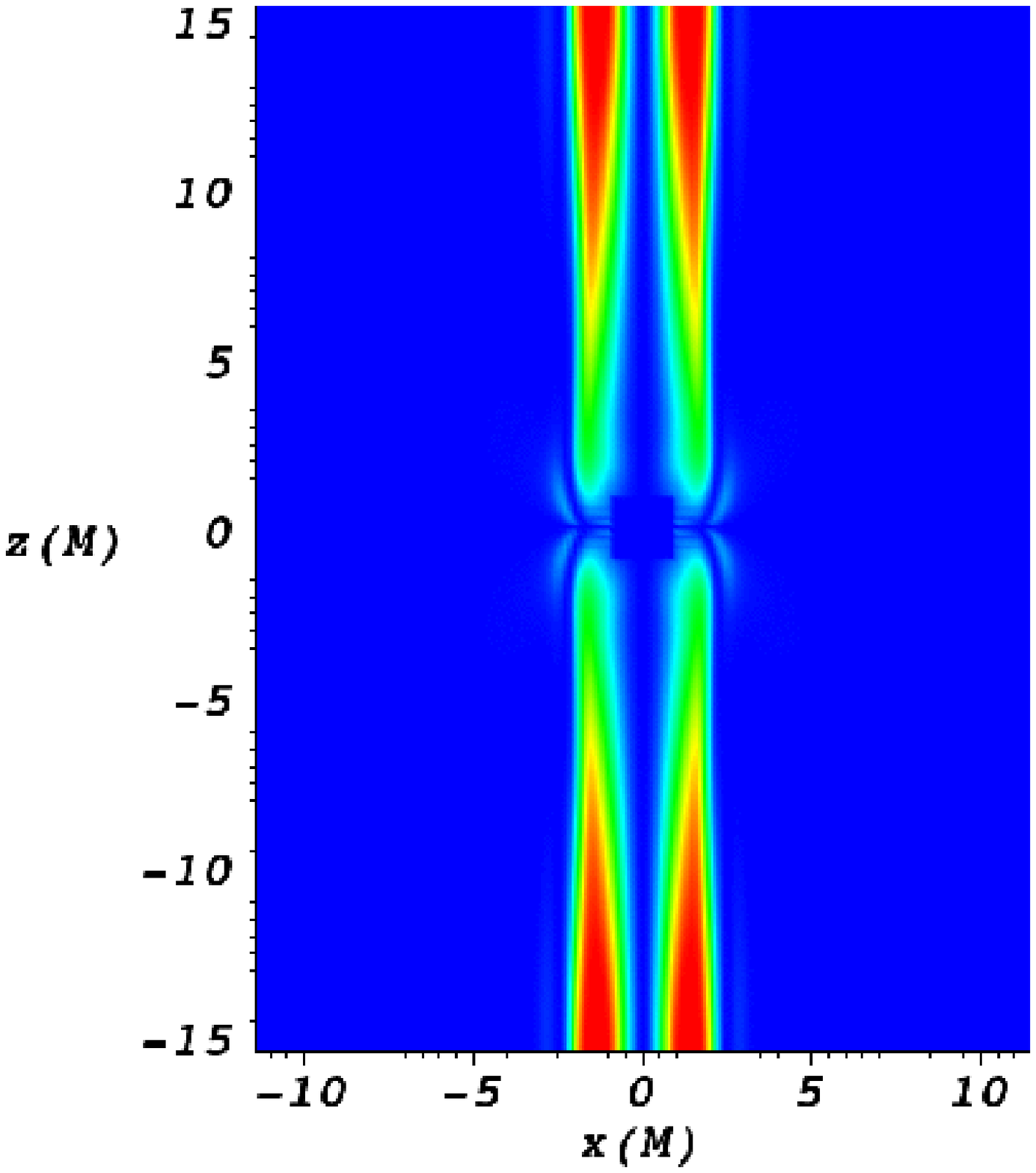, height=5.0cm,width=4.2cm}
\epsfig{file=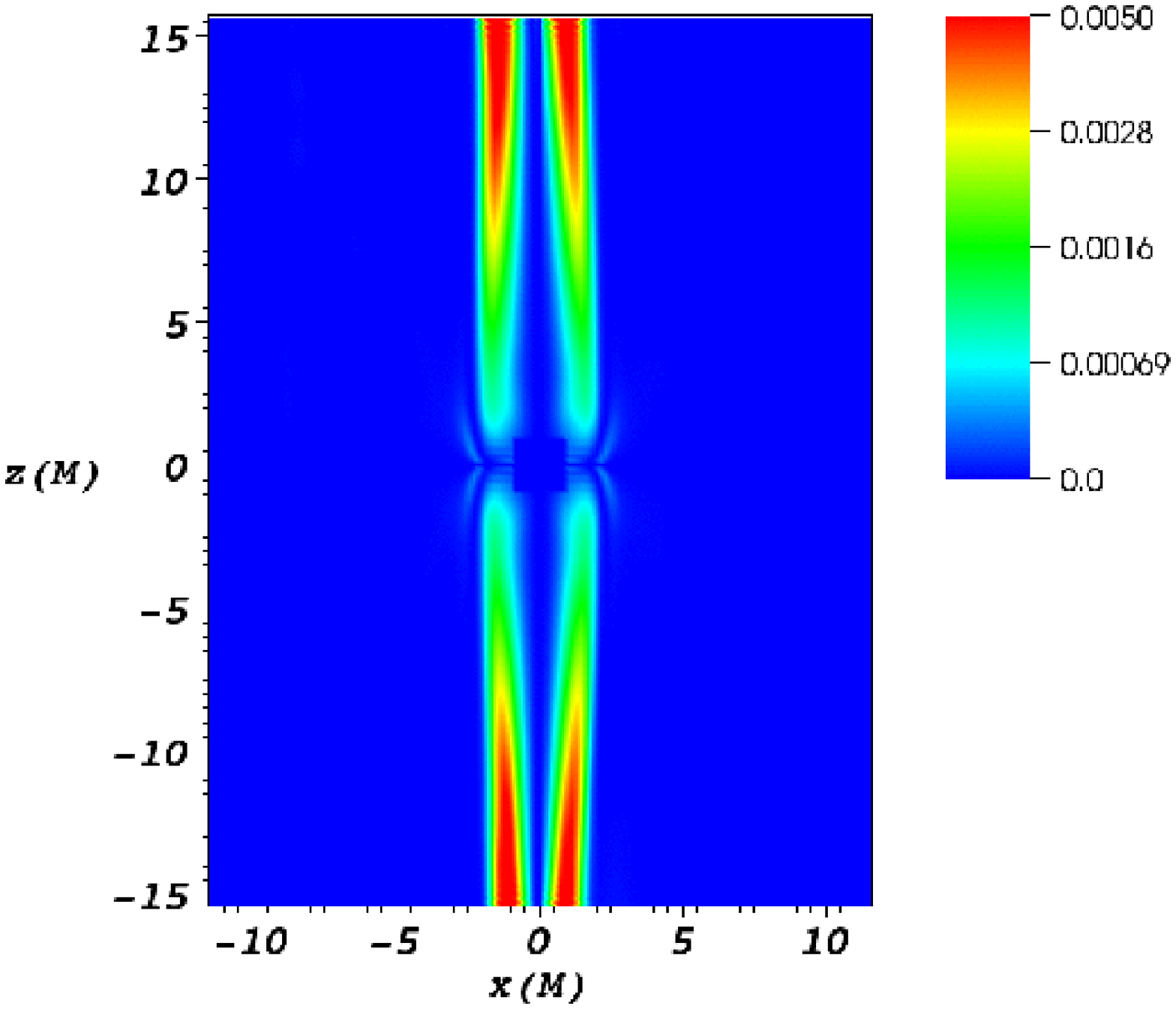, height=5.0cm,width=4.2cm}
\caption{Electromagnetic energy density flux for $R=0$ (left), $R=1/2$ (middle) 
and $R=1$ (right) at $t=320M$. }
\label{fig:jets}
\end{center}
\end{figure}

\begin{figure}
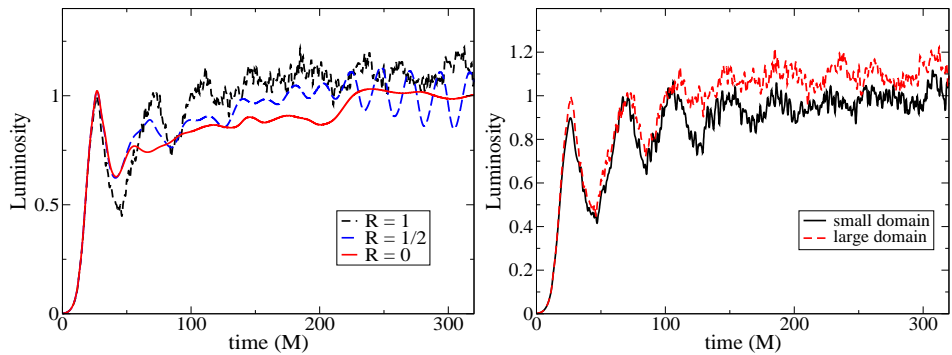

\begin{center}
\epsfig{file=luminosityR.eps, height=4.6cm,width=6.2cm}
\epsfig{file=domain.eps, height=4.6cm,width=6.2cm}
\caption{(Left) Luminosity for the three cases considered, normalized to
the asymptotic value of the $R=1$ case.
(Right) Luminosity for the two boundary locations considered in the case $R=1$,
normalized to the asymptotic value of the small domain simulation.}
\label{fig:domain}
\end{center}
\end{figure}

To examine the behaviour of the solution under these options we adopt a 
domain with $(x,y) \in [-12,12] M$ and $z = [-16,16] M$ covered with a uniform grid 
with grid-spacing given by $\Delta = 0.01 M$. We studied the three different cases
evolving them until a quasi-stationary regime is reached, comparing the
luminosity obtained for each case far away from the black hole. This luminosity 
is computed as the integral of the electromagnetic energy flux density (which can be seen for
the three cases studied in fig.~\ref{fig:jets})
on a solid angle of $15$ degrees along the z-axis (further details on this computation
are described in~\cite{2010arXiv1007.1198P}) and
shown in figure~\ref{fig:domain} (left plot), after a convenient renormalization with
the asymptotic value of the $R=1$ case.
After a transitory initial behaviour, the obtained luminosities for the different values
of the reflection coefficient $R$ are qualitatively the same in structure and the emitted
power within similar ranges. It is interesting to notice the energy density flux reaches a rather constant
asymptotic configuration in the $R=0$ case, while relatively small amplitude
cylindrical oscillations are observed for $0<R \le 1$. Nevertheless, the jet structure
and average luminosities obtained are comparable in all cases.
Thus, within typical astrophysical uncertainties it is clear the BZ mechanism
is not affected when considering significantly different boundary conditions. 

To ensure the boundary location does not affect the above observations, 
we have studied the jet's luminosity for two different boundary placements along the
jet's direction; the original domain (small) with $z = [-16,16] M$ and another one
(large) with $z = [-32,32] M$. Figure ~\ref{fig:domain} (right plot) illustrates the obtained values
for the case with $R=1$ normalized to the asymptotic value of the small domain simulation,
showing again that at late times the values obtained are quite similar.

\section{Final Comments}
This work illustrates the robustness of the BZ mechanism
to generate a collimated Poynting flux of energy which is 
largely independent of the boundary conditions adopted. Consequently,
while the load's characteristics are essentially unknown, our studies
indicate the jet resulting from the plasma's ability to extract
energy from the black hole's vicinity is robust.  

\section{Acknowledgments}
We would like to thank S. Phinney for pointing up this gap
in the understanding of the BZ mechanism and C. Thompson for
stimulating discussions on this problem.  
C.B. and C.P. thank FaMAF for its hospitality while parts
of this work were completed. 
This work has been jointly supported by European Union
FEDER funds and the Spanish Ministry of Science
and Education (projects FPA2008-04761-E and CSD2007-
00042); by NSF grant PHY-0803629; NSERC through a Discovery Grant
and the Canadian Institute for Advanced Research.
Research at Perimeter Institute is
supported through Industry Canada and by the Province of Ontario
through the Ministry of Research \& Innovation.  Computations were
performed at Scinet and Teragrid.

\section*{References}


\end{document}